\documentclass[12pt]{article}
\usepackage{graphicx}
\begin{document}

\centerline{\bf Opinion Dynamics with Hopfield Neural Networks}

\bigskip
Dietrich Stauffer*

\medskip
Condensed Matter Physics, Weizmann Institute of Science  

76100 Rehovot, Israel
\bigskip

Przemys{\l}aw A. Grabowicz and Janusz A. Ho{\l}yst

\medskip
Faculty of Physics, Warsaw University of Technology

Koszykowa 75, PL-00--662 Warsaw, Poland

\bigskip
*Visiting from Institute for Theoretical Physics, Cologne University,

D-50923 Cologne, Euroland

\bigskip
\centerline {Abstract}

{\small 
In Hopfield neural networks with up to $10^8$ nodes we store two patterns
through Hebb couplings. Then we start with a third random pattern which is 
supposed to evolve into one of the two stored patterns, simulating the 
cognitive process of associative memory leading to one of two possible opinions.
With probability $p$ each neuron independently, instead of following the 
Hopfield rule, takes over the corresponding value of another network, thus
simulating how different people can convince each other. A consensus is achieved
for high $p$.
}
\bigskip

One of the well-studied fields in sociophysics [1] is opinion 
dynamics. If one can chose between only two possible opinions, the human being 
is reduced to a binary variable $\pm 1$, just as in an Ising magnet the spin
is either up or down without consideration of the atomic structure leading
to that spin. However, our cognitive processes happen in the human brain, which
consists of $\sim 10^{11}$ neurons with $\sim 10^4$ connections each. 

A simple brain model is the Hopfield neural network: Each neuron $i$ is a 
binary variable $S_i = \pm 1$ connected to all other neurons $I$ through 
synaptic couplings $J_{iI}$. The neuron may switch its state following the 
sign of the sum of all interactions:

$$ S_i = {\rm sign} \sum_I J_{iI} S_I \quad . \eqno(1) $$

Usually the cognitive process studied in this model is the associative memory:
Starting from some unclear and rather random $S_i$, the above algorithm should 
eventually lead to one of the many previously stored patterns $\xi_i^{\mu}$,
where $\mu = 1,2 \dots$ counts the stored patterns. These patters are stored
by the Hebb rule

$$ J_{iI} = \sum_{\mu} \xi_i^{\mu}\xi_I^{\mu}  \quad .  \eqno(2) $$

We store two random patterns $\mu = 1$ and 2 on a $L \times L$ square lattice,
corresponding to two possible opinions, and start with random $S_i = \pm 1,
\; i=1,2\dots,L^2$. A complete and memory saving (Penna-Oliveira trick) 
computer program is given in [2] and is the starting point of 
the new simulations here. The single $S_i$ now should not be interpreted as
a single neuron but as part of our thinking leading us towards making
a decision.

Our new element is the simultaneous simulation of $k$ different Hopfield 
networks, coupled to each other. Thus at each iteration each neuron 
independently deviates with probability $p$ from Eq(1) and instead takes the
value of the corresponding neuron (same $i$) from a randomly selected other 
network. Each of the $k$ different people then can learn from the others or
convince them. In this way some cognitive process is simulated, instead of 
simple flips of opinions $\pm 1$.

\begin{figure}[hbt]
\begin{center}
\includegraphics[angle=-90,scale=0.3]{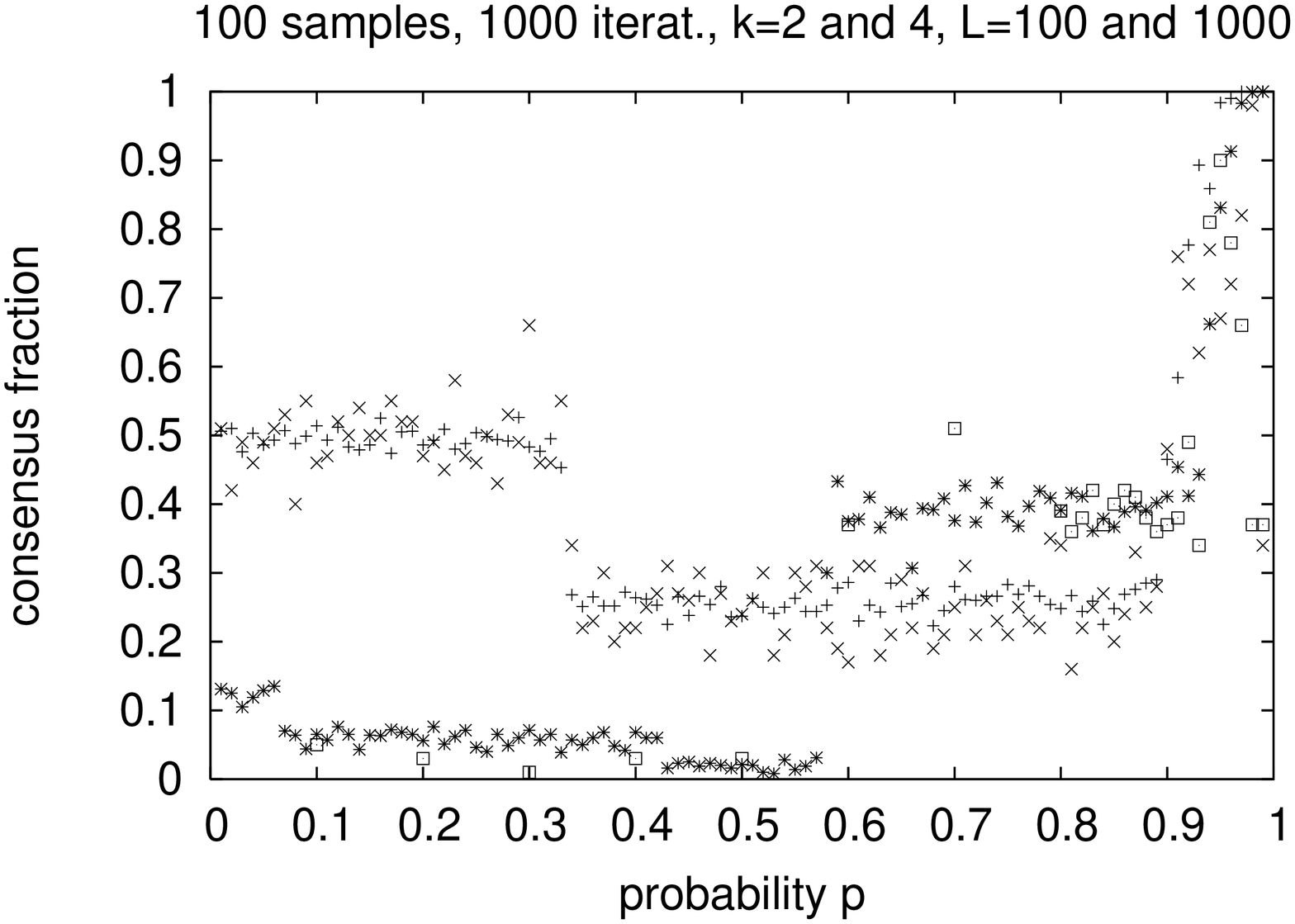}
\includegraphics[angle=-90,scale=0.3]{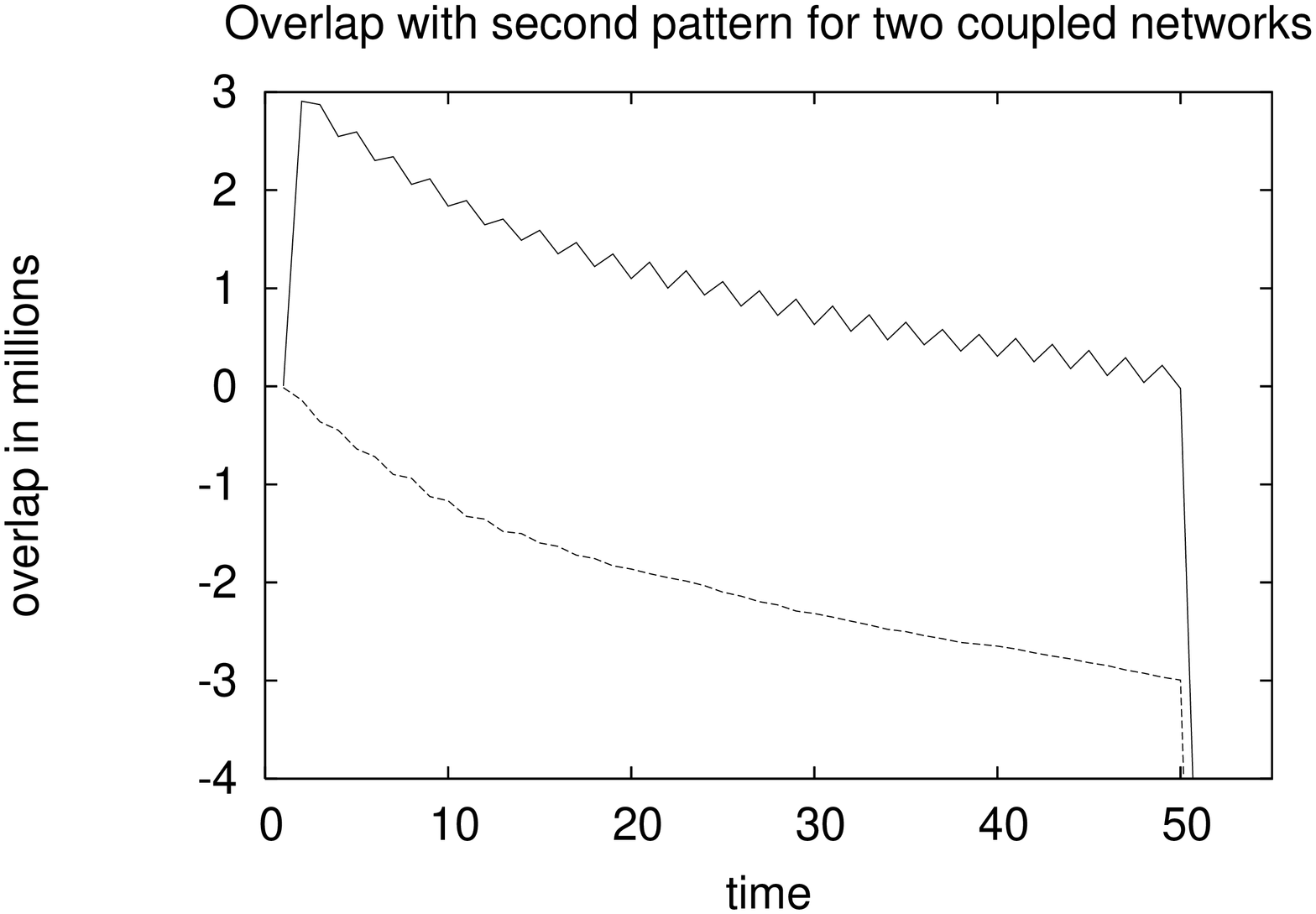}
\end{center}
\caption{
Top: Fraction of samples leading to agreement for $k = 2$ (+,x) and 4 (*, 
squares) with linear pattern dimension $L = 100$ (+,*) and 1000 (x, squares).
1000 iterations were used. Bottom: Neural dynamics of two people with $p = 0.97,
\; L = 10,000 \; (10^8$ neurons). We show the overlap $\sum_i S_i \xi_i^2$
with the finally winning second opinion, which reaches --95 million after
100 iterations.
}
\end{figure}

\begin{figure}[hbt]
\begin{center}
\includegraphics[angle=-90,scale=0.29]{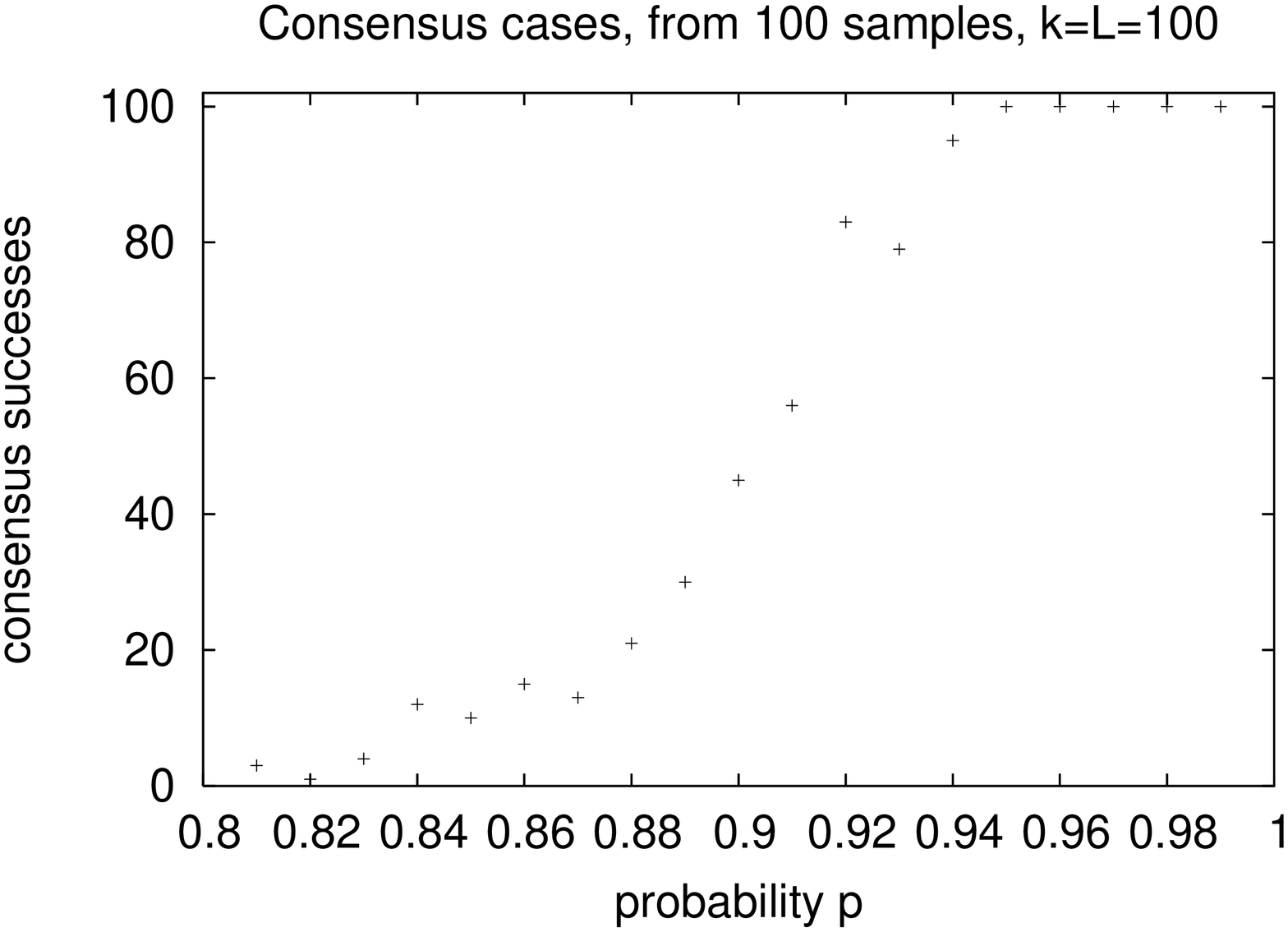}
\includegraphics[angle=-90,scale=0.29]{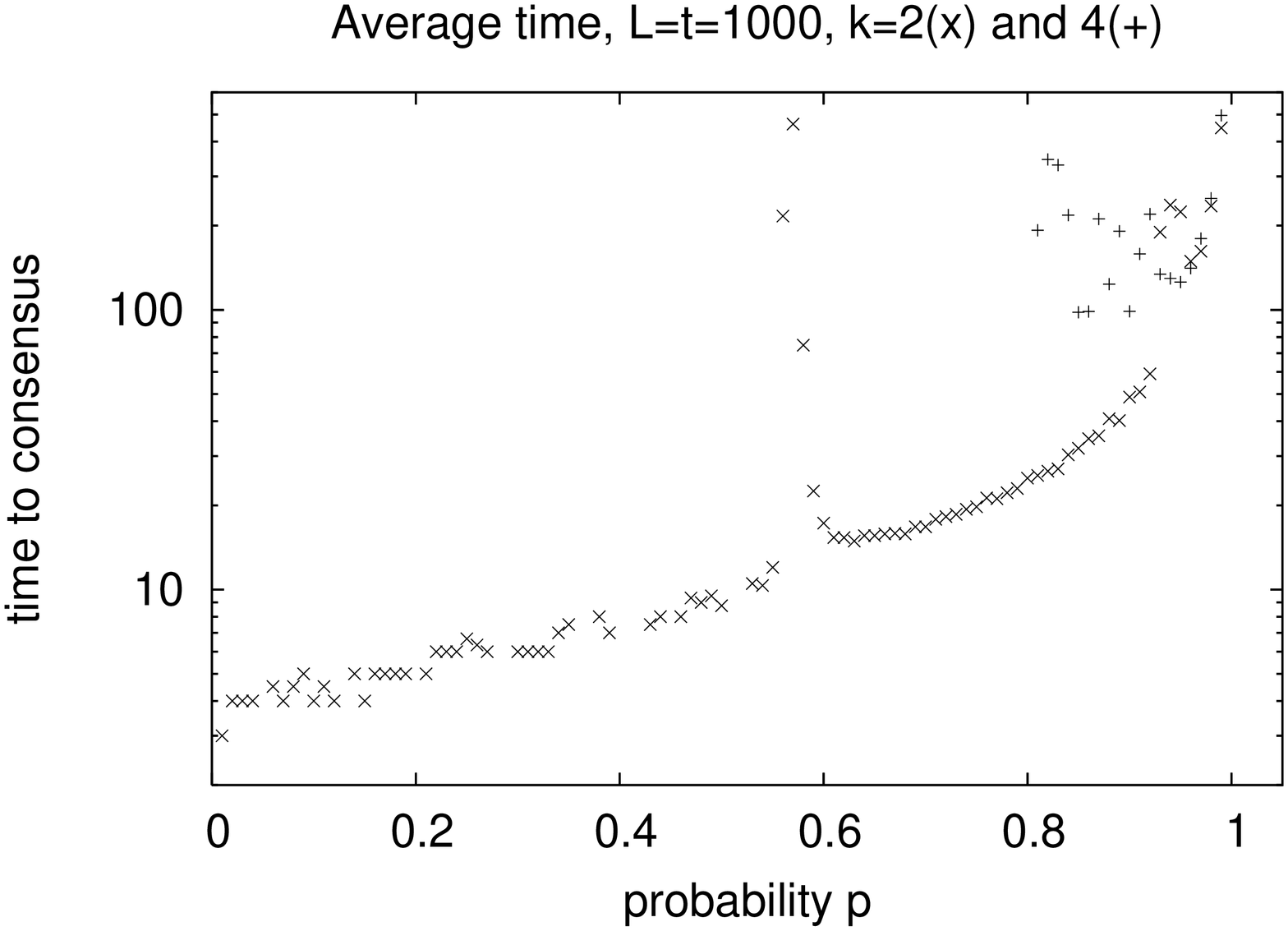}
\includegraphics[angle=-90,scale=0.29]{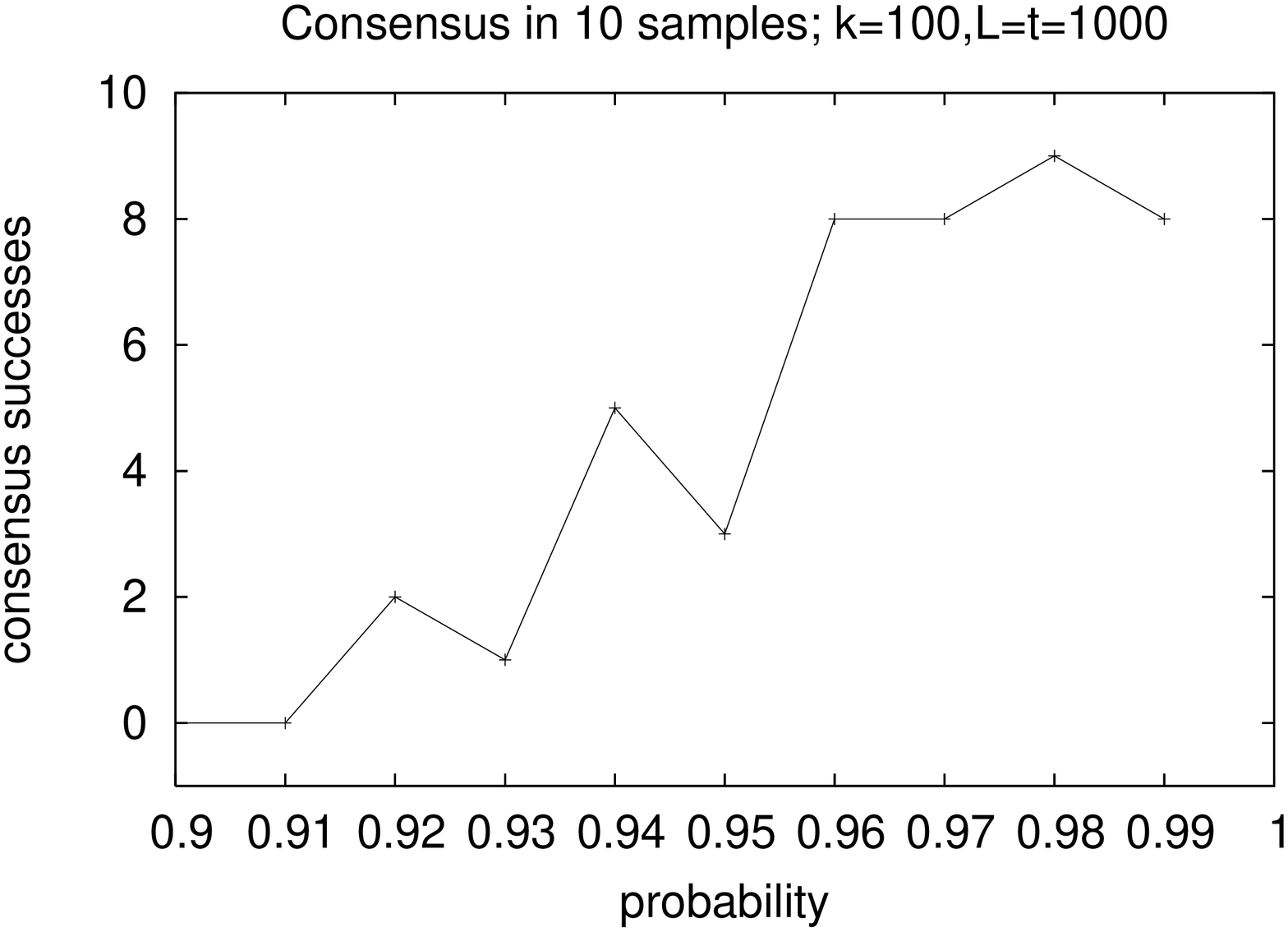}
\end{center}
\caption{
Top: Number of samples (from 100) leading to agreement for $L = k = 100, \; 
t=1000$. Centre: Average number of iterations needed for consensus of top part, 
ignoring the samples where no consensus was reached within 1000 iterations.
Bottom: As top, but for hundred times more neurons.
}
\end{figure}

\begin{figure}[hbt]
\begin{center}
\includegraphics[angle=-90,scale=0.3]{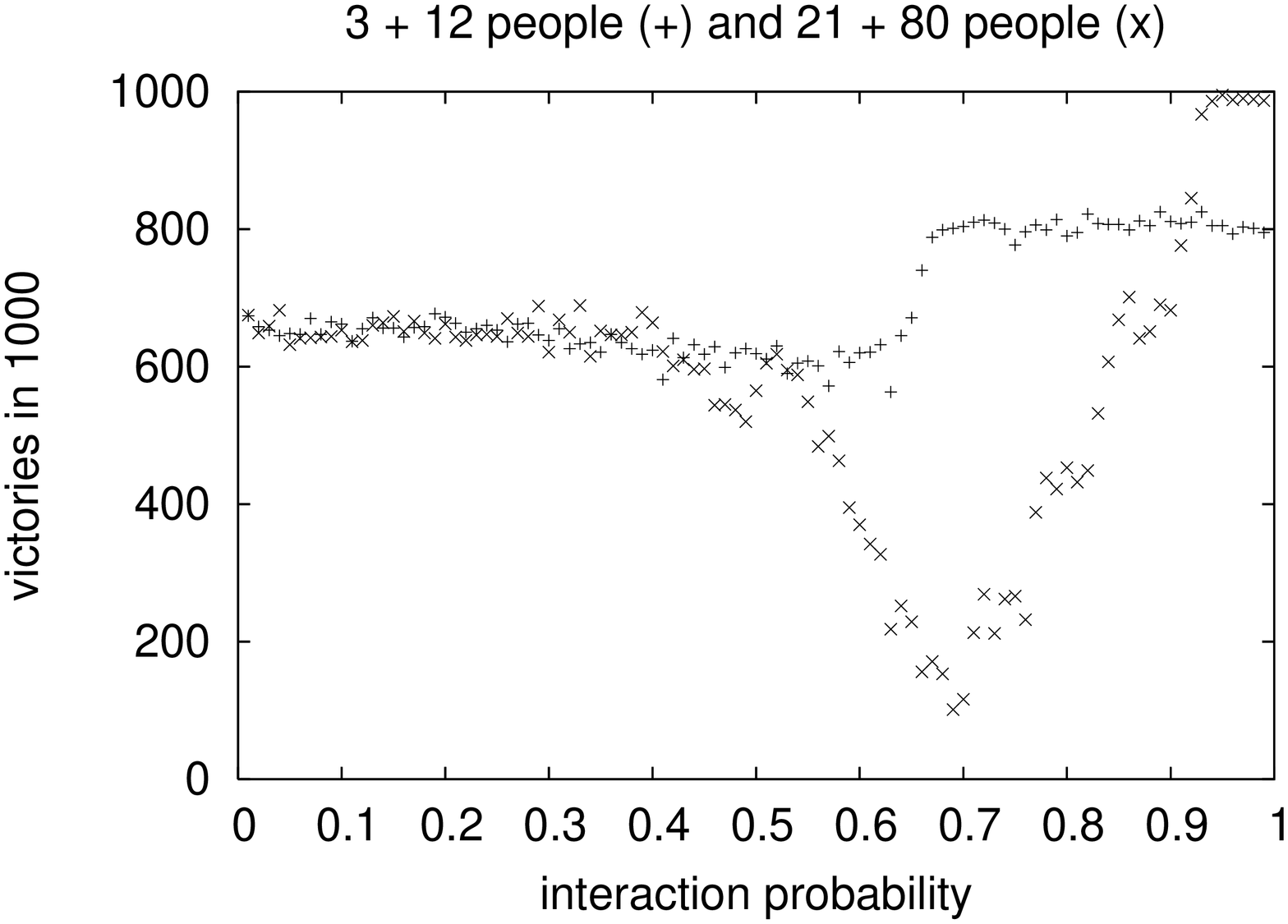}
\includegraphics[angle=-90,scale=0.3]{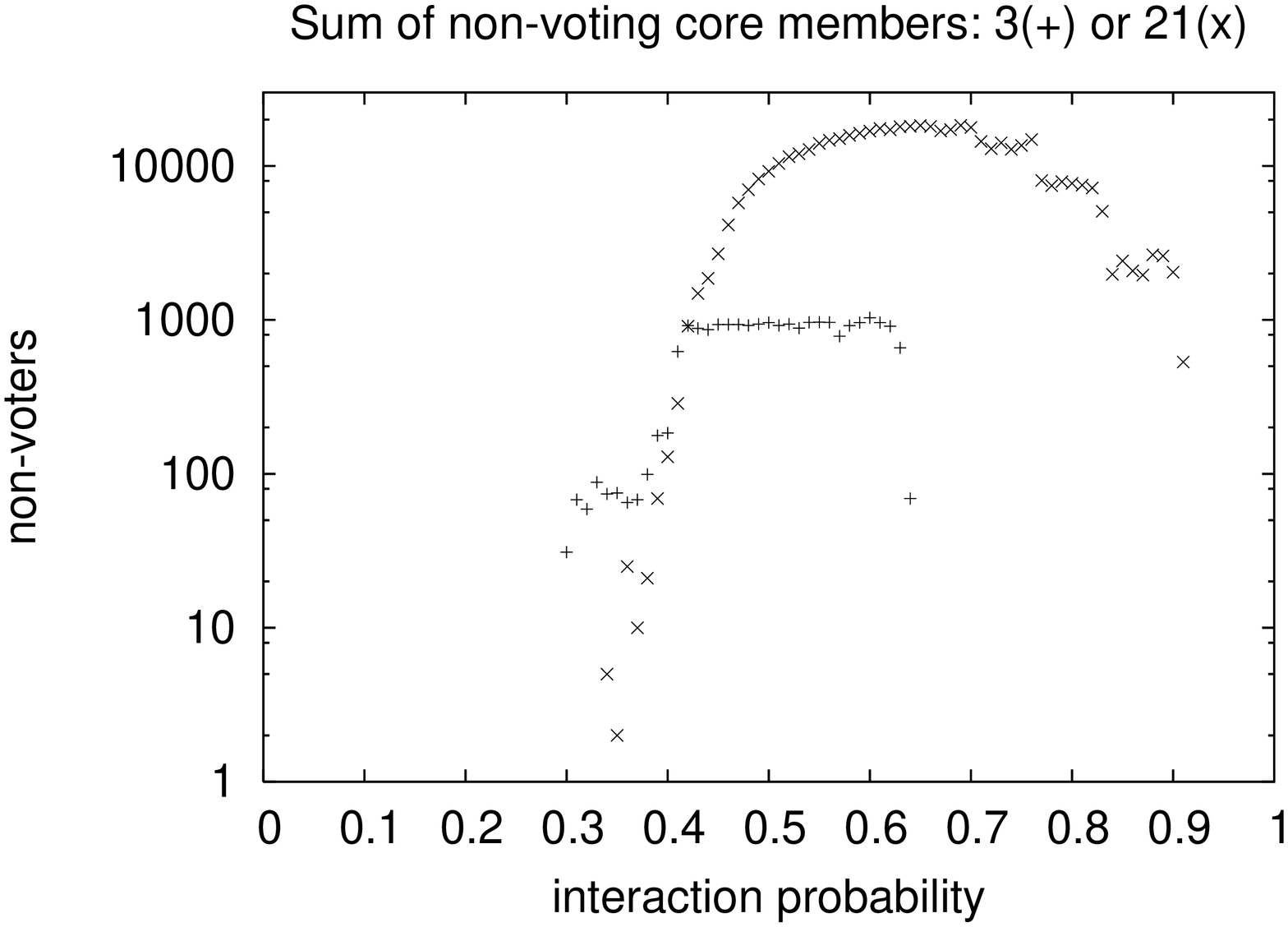}
\end{center}
\caption{Influence of an interacting core of 3 or 21 members on the majority of 
all $k = 15$ or 101 voters, respectively, summing over 1000 samples. The top
shows victories, defined as samples with the overall majority agreeing with the 
core majority; the lower case shows the summed number of core members who were 
unable to converge on a pattern. At least 75 \% agreement means converegence.
}
\end{figure}

We found that for $L = 100$ (thousand samples), 1000 (hundred samples) and
10,000 (one sample) that at low $p$ the opinions end up randomly (for example
agreement in half the cases if $k = 2$), at intermediate $p$ less often 
agreement was found, and for $p$ close to one agreement was found more often,
including all thousand cases for $k = 2, \; L = 100, \; p \ge 0.97$; see
Fig.1. For $L = k = 100$ no random agreement is possible and only high $p$
achieved consensus; see Fig.2. More iterations are needed at high $p$ to find
agreement.

We see that for small $k$ an accidental consensus is possible for small 
coupling $p$. At intermediate $p$ the continuous discussions may prevent
the participants to reach any of the two opinions; then even accidental 
agreement is impossible. For very large $p$ consensus can be reached always.
If the number $k$ of people is large, then accidental consensus is impossible,
and only large $p$ produce consensus.

(As in single Hopfield models, $k=1$, the agreement with the stored pattern is 
sometimes but not always complete; we counted agreement if 75 \% of the stored 
pattern was recovered correctly. Sometimes, however, the system cannot decide 
which of the two stored patterns it should converge to, and remains blocked
apart from minor fluctuations. And as in single Hopfield models, "agreement" 
can also mean the convergence towards the complementary pattern, leading 
for all $i$ to $S_i = +1$ where $\xi_i = -1$ and vice versa; see Fig.1b.)

If $k=2$ (two networks A and B) and if $a$ is the fraction of $S_i$ having
the correct value of a stored pattern or its complement, then we define as a 
criterion for ``agreement'' that $a \ge a_c$ where the threshold value $a_c$
was taken as 
$a_c = 0.75$ in our simulations. Without interactions, $p=0$, after a few 
iterations we have $a=1$ for both networks. Let us assume that network B is the 
opposite (complement) of network A. With probability $p \ll 1$ at each 
interation, each $S_i$  of A gets a ``wrong'' value from network B, and thus 
averaged over all $i$: $a=1-p$. A more detailed evaluation gives $p = (1-a)/a$ 
which means $p_c=1/3$ for our numerical choice $a_c=3/4$, in agreement with 
the jump observed in Fig.1a. The more accurate the required agreement is,
the smaller must the probability $p$ be to lead to a random consensus, i.e
high $a_c$ require low $p_c$. 
(In the cases where network B is not the complement of network A, the agreements
$a$ of the networks with the patterns are less vulnerable to disruption
and thus the threshold value for $p$ is higher. $p=(1-a)/a$ thus is the lowest 
$p$ value at which the consensus starts to be non-random.)

Decision-making committees are often dominated by a smaller core of interacting
people. We simulate this ``old-boy network'' by assuming that only the core
members interact with each other in the above way, while the remaining people
do not interact with anybody and thus arrive at random decisions. A victory
for the core is defined as a case where the majority vote of all agrees with
the majority vote of only the core. Fig.3 shows random victories at low
$p$ and nearly complete victories for very high $p$. But at intermediate 
$p$ again many core members only discuss instead of arriving at a decision,
and often the core does not cast a single vote. (It does not matter if we 
give the core 100 or 1000 iterations for deliberations.) 

This feasibility study is a neural network generalisation of the voter model,
where everybody can take over the opinion of a randomly selected neighbour [3],
somewhat similar to the Axelrod model [3].
Another application could be the Sznajd model of convincing [4,2].

The work was partially funded by the European Commisssion Project CREEN 
FP6-2003-NEST-Path-012864, and by Polish Ministry of Science and Higher 
Education under Grant 134/E-365/6, PR UE/DIE 239/2005-2007.

{\small
\begin{verbatim}

      program neuropin
c     ixi(i,m,j): neuron i=1...L*L;pattern mu=1,2;network j=1...k. 
      parameter(L=100 ,n=L*L,k=  4)
      dimension ixi(n,2,k),is(n,k),m(2,k),ifixed(k)
      logical same,samp
      byte ixi,is,one
      data iseed/1/,nrun/100/,max/1000/,one/1/
      fact=0.5/2147483647
      ibm=2*iseed-1
      print *, L,k,iseed,nrun,max
      do 10 ipr= 1,88,1 
      p=0.01*ipr
      ip=(2*p-1)*2147483648.0d0
      icnt=0
      do 2 mu=1,2
      do 2 i=1,n
        ixi(i,mu,1)=-one
        ibm=ibm*16807
 2      if(ibm.gt.0) ixi(i,mu,1)=one
      do 1 j=2,k
      do 1 mu=1,2
      do 1 i=1,n
 1      ixi(i,mu,j)=ixi(i,mu,1)
      icount=0
      do 8 irun=1,nrun
      call flush(6)
      do 3 j=1,k
      do 3 i=1,n
        ibm=ibm*16807
        is(i,j)=one
 3      if(ibm.lt.0) is(i,j)=-one
      if(L.eq.38) print 100, (ixi(i,1,1),i=1,n)
c     initialisation of 2 fixed patterns + 1 variable pattern 
      do 4 itime=1,max
      do 5 j=1,k
      do 5 mu=1,2
        m(mu,j)=0
        do 5 i=1,n
 5        m(mu,j)=m(mu,j)+is(i,j)*ixi(i,mu,j)
      do 6 j=1,k
      ifixed(j)=0
      do 6 i=1,n
        isold=is(i,j)
        ifield=ixi(i,1,j)*m(1,j)+ixi(i,2,j)*m(2,j)
        is(i,j)=one
        if(ifield.lt.0) is(i,j)=-one
        ibm=ibm*16807
        if(ibm.gt.ip) goto 6
 9      ibm=ibm*65539
        jj=1+(ibm*fact+0.5)*k 
        if(jj.le.0.or.jj.gt.k.or.jj.eq.j) goto 9
        is(i,j)=is(i,jj)
 6      ifixed(j)=ifixed(j)+isold*is(i,j)
      same=.true.
      samp=.true.
      do 12 j=1,k
        same=same.and.(ifixed(j).eq.n) 
 12     samp=samp.and.(iabs(m(1,j)).eq.n.or.iabs(m(2,j)).eq.n)
      if(L.gt.5000) print *, irun,itime,m,ifixed,icnt,same,samp
      if(same.and.samp) icount=icount+1
      if(same.and.samp) goto 13
 4    continue
      if(L.eq.38) print 100, ((is(i,j),i=1,n),j=1,k)
 13   samp=.true.
      do 11 j=2,k
      do 11 jj=1,j-1
      same=.false.
      do 7 mu=1,2
        x1=iabs(m(mu,j))
        x2=iabs(m(mu,jj)) 
 7      same=(x1+x2.gt.n.and.abs(1.00-x1/x2).lt.0.1).or.same
c     print *, j, jj, samp, same
 11   samp=samp.and.same
      if(samp) icnt=icnt+1
      print *, irun,itime,icnt,samp
      call flush(6)
 8    continue
 10   print *,p,icnt,icount
 100  format(1x,38i2)
      end
\end{verbatim}
}

\begin{thebibliography}{99}

\bibitem{c} C. Castellano, S. Fortunato and V. Loreto, subm. to Rev.
Mod. Phys., arXiv: 0710.3256

\bibitem{n} D. Stauffer, S. Moss de Oliveira, P.M.C. de Oliveira and J.S. 
S\'a Martins, {\it Biology, Sociology, Geology by Computational Physicists},
Elsevier, Amsterdam 2006

\bibitem{mallorca} M. San Miguel, V.M. Egu\'{\i}luz and R. Toral, Comp. Sci. 
Engin. 7, 67 (Nov/Dec/ 2005)

\bibitem{sznajd} K. Sznajd-Weron and J. Sznajd, Int. J. Mod. Phys. C 11, 1157
(2000). 

\end{thebibliography}
\end{document}